\documentclass[prl,preprint,floatfix,showpacs,superscriptaddress]{revtex4}
\usepackage{amsmath}
\usepackage{graphicx}% Include figure files

\begin{document}

\title{Electronic state of a doped Mott-Hubbard insulator at 
finite temperatures studied using the dynamical mean-field theory}

\author{A. Camjayi}
\affiliation{Departamento de F\'{\i}sica, FCEN, Universidad de Buenos Aires,
Ciudad Universitaria Pab.I, Buenos Aires 1428, Argentina.}
\author{R. Chitra}
\affiliation{Laboratoire de Physique Theorique des Liquides, UMR 7600, 
Universite de Pierre et Marie Curie, Jussieu, Paris-75005, France}
\author{M. J. Rozenberg}
\affiliation{Laboratoire de Physique des Solides, CNRS-UMR8502, Universite de Paris-Sud,
Orsay 91405, France.}
\affiliation{Departamento de F\'{\i}sica, FCEN, Universidad de Buenos Aires,
Ciudad Universitaria Pab.I, Buenos Aires 1428, Argentina.}
 
\date{today}
\begin{abstract}
 
We study the electronic state of the doped
Mott-Hubbard insulator within Dynamical Mean Field Theory.
The evolution of the finite temperature spectral functions as a function
of doping show large redistributions of
spectral weight in both antiferromagnetic and paramagnetic phases.
In particular, a metallic antiferromagnetic state is obtained with a low
frequency Slater-splitted quasiparticle peak coexisting with Hubbard bands.
In the high temperature paramagnetic metallic phase, upon reducing 
doping, the system has a crossover through a ``bad metal'' state characterized
by an anomalous shift of the quasiparticle peak away from the Fermi energy. 
We find that the {\it charge} compressibility of the antiferromagnetic metal 
is dramatically enhanced upon approaching the second order N\'eel line.

%We solve the DMFT equations for the Hubbard model across
%the finite temperature metal-insulator transition as a function of
%doping. We follow the evolution of the spectral functions from
%the strongly correlated paramagnetic metallic state at finite doping
%towards the half-filled Mott insulator. At higher temperatures the
%insulator is paramagnetic and at lower $T$, is an antiferromagnet.
%In the high temperature paramagnetic phase, upon reducing (hole) doping in the metallic phase,
%we observe an anomalous
%loss of spectral weight of
%the quasiparticle peak at the Fermi energy. 
%As we enter the antiferromagnetic phase at lower temperatures,
% the system
%remains metallic for a certain doping range and we observe
%a Stoner-like splitting of the quasiparticle peak which coexists with the polarized high frequency 
%Hubbard bands.
%In addition, the compressibility shows a strong anomaly at the magnetic
%boundary that eventually diverges at a finite low temperature. We discuss the possible
%experimental implications of our results.
%These anomalies  should be observed in photoemission, STM and ultrasound velocity attenuation
%experiments of strongly correlated compounds with 
%antiferromagnetic insulating phases.
 
\end{abstract}


\maketitle
%{\it Introduction.}
The understanding of metal-insulator transitions is a 
central problem of condensed
matter physics. One of the most intriguing is the Mott
transition, where the itinerant electrons of a metal localize
due to strong correlation effects stemming from Coulomb repulsion
\cite{mott}. In a Mott state, though the electrons become localized, their spin (and
possibly orbital) degrees of freedom remain unquenched. This leads to
a large entropy and consequently,  Mott
insulators  tend to order magnetically (and/or orbitally)  at sufficiently low
temperatures \cite{ift}.
Examples of strongly correlated systems where the Mott phenomenon occur
include materials with
partially filled $d$ and $f$ orbitals.
The Mott transition can be driven by pressure, temperature or doping and
exotic physics such as high temperature superconductivity, colossal 
magnetoresistance and heavy fermion states are usually
found in the vicinity of  Mott insulating states \cite{ift}.

A minimal model that captures the physics of the Mott transition 
is the Hubbard model.
The development of the Dynamical 
Mean Field Theory (DMFT)\cite{rmp} enabled progress in
the theoretical understanding of the
 Mott-Hubbard transition in the limit of large lattice coordination.
Moreover, the combination of DMFT with ab-initio methods   is a promising new tool
to study strongly correlated real materials \cite{kotliar_vollhardt} which are 
often described by   complex multi-band  hamiltonians. In the light of these developments,
a clear and reliable solution of basic  models like the Hubbard model is  an
indispensable stepping stone.
Much of the work  has been centered on the study
of the {\em frustrated} Hubbard model which has no 
antiferromagnetic (AF) order and shows a first order
Mott transition line that ends in a 
finite temperature second order critical point 
\cite{rmp,prls, limelette}.
The original  Hubbard model, however, has 
a Mott insulating
phase at low temperature with N\'eel  order. 
This AF order can be destroyed by increasing temperature,
leading to a  paramagnetic (PM) Mott insulator; or by doping, leading
to a strongly correlated PM  metal \cite{rmp}.
The dramatic changes of the electronic state of the system as it is
doped away from the  Mott state to the correlated metallic state are of great current
interest. 
For example, cuprates first go through an intriguing  pseudogap
state \cite{rmp_lee} followed  by a strange  metallic 
one with anomalous properties,  to end
in a  strongly correlated Fermi liquid \cite{rmp_lee,ando,shen}. Other compounds
like the   titanates  LaTiO$_3$ and CaTiO$_3$ and  the vanadate 
V$_2$O$_3$, also  
exhibit anomalous behaviors with doping, such as a divergent effective mass
and large transfers of spectral weight \cite{ift}.

Here, we focus on  
the  finite temperature doping driven  metal-insulator transition in the 
original  Hubbard model, i.e. {\em without frustration}. 
We present a detailed solution of the model
with  particular emphasis on the
systematic changes of the density of states (DOS). A highly correlated AF metal phase
is found and 
contrary to naive expectations,  
the onset of magnetism has a dramatic effect on the  electronic compressibility. We find
that the  high
temperature PM metallic phase also shows novel behavior at very low doping.
These   properties are    clearly  relevant  for interpretations
of recent experiments in photoemission where the dependence of the doping on 
chemical potential was precisely measured \cite{shen}, and also for
STM spectroscopies and ultra-sound velocity attenuation amongst others.

The Hubbard model reads,
\begin{equation}
 H = t\sum_{\langle ij\rangle \sigma }[c^+_{i\sigma} c_{j\sigma} + h.c.] 
- \mu\sum_{i\sigma} 
%c^+_{i\sigma} c_{i\sigma} + U\sum_i (n_{i\uparrow}- \frac12)(n_{i\downarrow}-\frac12)
n_{i\sigma} + U\sum_i (n_{i\uparrow}- \frac12)(n_{i\downarrow}-\frac12)
\end{equation}
where the hopping $t$ is between nearest neighbors, $c^+_{i\sigma}$ creates
a particle with spin $\sigma$ at site $i$, $U$ is the Coulomb repulsion
and $\mu$ is the chemical potential. 
In the limit of
large lattice coordination, the above model can be exactly mapped onto a single
impurity Anderson model with a  supplementary  self-consistency 
condition for the hybridization function \cite{rmp}.
For simplicity, we consider a Bethe lattice which  is  bipartite and has  
a semicircular density of states, and therefore, well adapted  
to the study of commensurate magnetism.
 The results are expected to be 
 qualitatively similar on  a hypercubic lattice.

The magnetic phase diagram of the model on a hypercubic lattice as a function of
doping
was obtained in [\onlinecite{freericks}] by computing the spin suceptibility in the
PM phase.
In that work, it was found that the AF instability is towards a commensurate
state for most of the parameter space. The instability towards incommensurate
states was found at very low temperatures and for a small range of doping.
In contrast to Ref.[\onlinecite{freericks}], in this Letter, we 
consider N\'eel  AF order and we explicitly
solve for the DMFT equations in the broken symmetry phase. 
This allows us to directly obtain the spin dependent local Green's function and
hence the DOS. 
Our work does not touch upon the question of phase separation at very low temperatures since we work in
temperature ranges and strong coupling where stable commensurate spin order is
expected to occur.  The DMFT is not the optimal method to study inhomogeneous phases and other
exotic orderings. [\onlinecite{zitzler}] 

When  AF  N\'eel long range order is allowed
on the original lattice, the DMFT self-consistency condition 
for the associated Anderson impurity model reads \cite{rmp},
\begin{equation} \label{selfcon}
{\cal G}^{-1}_{o\sigma}(i\omega_n) =
i\omega_n + \mu - t^2 G_{-\sigma}(i\omega_n)
\end{equation}
${\cal G}_{o\sigma}$ and $G_{\sigma}$ are the Green's functions of the
associated impurity. 
$t^2 G_{-\sigma}(i\omega_n)$ plays the role of the
 impurity hybridization $\Delta_{\sigma}(i\omega_n)$.
When the self-consistency condition (\ref{selfcon}) is fulfilled, the impurity Green's
function $G_{\sigma}$ coincides with the local Green's function of the lattice \cite{rmp}.
We use the quantum Monte Carlo method of Hirsch and Fye \cite{hf} 
to obtain the exact solution (in the 
statistical sense) of the model in the strong correlation regime,
 which is beyond the scope of analytic methods.
In order to obtain the DOS, one has to analytically continue
the Monte Carlo data to the real axis using a maximum entropy
method \cite{gubernatis}. 
Due to the critical fluctuations  near the AF/PM boundary,
expensive simulations are required to obtain numerically reliable
results;  we, therefore, focus on a single
value of the Coulomb repulsion $U=3.125$ and the unit of energy
is set by the half-bandwidth $D=2t=1$.
As expected, we find that at
half-filling ($\mu = 0$) the system is an AF Mott insulator with a large gap
($\sim D$)  
below a N\'eel temperature of $T_N \simeq 0.1$, and a PM 
Mott insulator above. 

We begin with the discussion of the
evolution of the DOS as a function doping.
We focus on two temperatures which are  representative of
the two characteristic evolutions of the DOS:
$T < T_N$ allows  us to trace the evolution of the  
system from an  AF Mott insulator at half-filling  to  a  strongly correlated PM metal
 as it traverses  a second order AF/PM transition line; 
$T > T_N $  permits us to  
study the {\em crossover} from the PM Mott insulator to 
the strongly correlated PM metal through a ``bad metal'' state. 

\begin{figure}
  \centerline{\includegraphics[scale=0.5,angle=270]{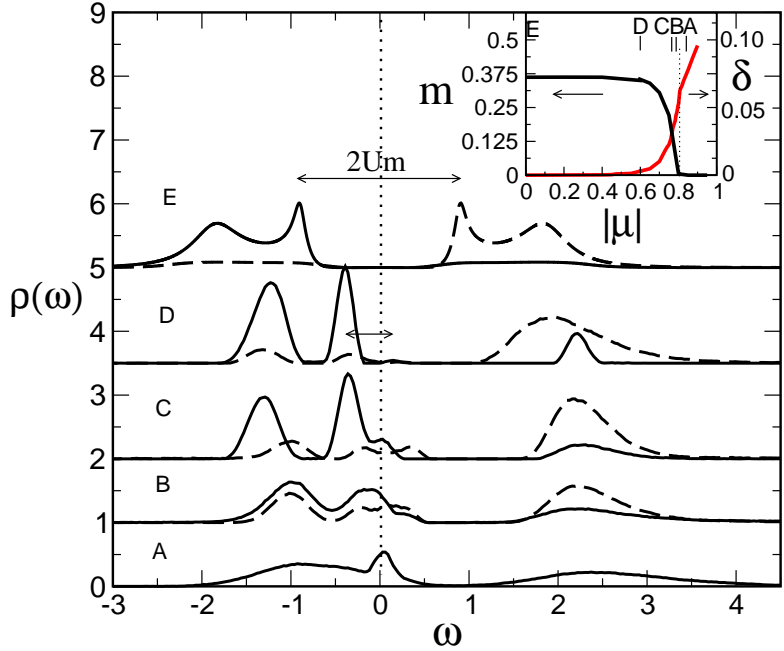}}
%\centerline{\includegraphics[angle=270]{fig1.ps}}
\caption{The local DOS $\rho(\omega)$ for $U=3.125$ and
$T=0.0625$ for  values of $-\mu= 0.84, 0.8, 0.777, 0.6, 0.0$ which correspond
to dopings $\delta= 0.076, 0.053, 0.037, 0.003, 0.0$ respectively
from bottom to top. Each line type corresponds to a spin projection.
The double arrow head line indicates the large Mott gap (E) and its
dramatic reduction upon infinitesimal doping (D).
The inset shows the staggered magnetization and the doping as a function
of $|\mu|$. The letters A to E indicate the values of $\mu$ that
corresponds to the spectra in the main panel. The dotted line indicates the
AF/PM boundary.
}
\label{fig1}
\end{figure}

In Fig.\ref{fig1}, we consider $T=0.0625 < T_N$ and show the DOS
for the two spin projections and different values of the chemical
potential $\mu$.
We consider the case of hole doping, therefore we take $\mu \leq 0$.
The top most spectra (E)  corresponds to the half-filled AF Mott insulator (AF-I) at $\mu=0$ and
displays up-down symmetry. This state has a rather large
staggered moment $m$ (inset) favored by the bipartite nature
of the lattice and the large value of $U/D$.
Expectedly,
the spectra shows Hubbard bands separated by a large gap of size $\approx 2Um$
(indicated in the figure by double head arrows).
The lower edges of the Hubbard bands are strongly enhanced 
due to the effective doubling of the unit cell in the N\'eel state. 
As the inset shows, the chemical potential can be varied within the 
large Mott gap ($|\mu| \stackrel{<}{\sim} 0.6$) without changing the filling
and 
the staggered magnetic moment $m$ displaying a plateau.
Accordingly, one finds that the DOS merely shows
a rigid shift in energy of the spectra E. When 
$\mu$ reaches the
Hubbard band edges, carriers are doped to the system and
dramatic changes in the spectral functions are observed.
They can be more easily understood by following the evolution of the
DOS from
 the high doping PM  end, i.e.,  from  the PM metallic  spectra A.
It is a
strongly correlated PM metallic state that shows a narrow quasiparticle peak 
at the Fermi energy flanked
by the  lower and upper Hubbard bands.
The quasiparticle peak carries a reduced spectral weight
and its small width defines the renormalized Fermi energy  
$\epsilon_F^*$.  
Due to its proximity to a 
Mott transition, the effective mass of this state is greatly enhanced
compared to the non-interacting value 
and one has a heavy Fermi liquid (PM-FL) \cite{rmp,prb}.

The spectra B to D show the dramatic changes of the DOS as the system
evolves between the two limit cases A and E just described.
Reducing $\delta$ from the PM metal brings the system back to 
the AF phase.
The localized (Hubbard bands) and itinerant (quasiparticle peak) aspects 
of the DOS that coexisted in the
PM metallic correlated state survive the AF transition.
However, as one enters into the AF state they evolve in a qualitatively 
different manner: while the incoherent Hubbard bands
become polarized and merely modify their relative spectral weight, the 
coherent quasiparticle peak reveals a Slater like splitting. 
The latter feature
can be understood in two complementary ways: one  is to recall that 
in terms of the associated impurity model,  
the central peak 
in the correlated PM metal (spectra A)
corresponds to a Kondo  resonance.
On the AF side, the  polarization of the hybridization
function $\Delta_\sigma(i\omega_n)$ ({\cal cf.}(\ref{selfcon}))
can be interpreted as an effective (frequency dependent) 
magnetic field at the impurity site.
This effective field   then induces a Zeeman splitting of the Kondo 
peak. 
Another  way to understand the behavior of the AF state
is to think of the quasiparticle peak as 
a band narrowing due an effective renormalization of hopping by the interaction,
similar to the Brinkman-Rice solution of the Hubbard model \cite{rmp,br}.
Then, the AF splitting of the quasiparticle band  can be simply
understood 
as a Slater-like bandstructure effect due to the doubling of the lattice constant
when N\'eel magnetic order sets in. 
The splitting of the quasiparticle peak is quite small and the DOS for
both up and down spin projections are always {\em metallic} (AF-M).
This observation is qualitatively consistent with the  
unexpected metallic behavior seen in large regions of
the AF phase in the cuprate  ${\rm La}_{2-x}\rm{Sr}_x\rm{CuO}_4$ 
and the insensitivity of the resistivity across the PM/AF transition
in clean and very low doped samples ($\sim 1\%$) reported in Ref.\cite{ando}. 
%Note that our results show that the staggered magnetic moment
%has contributions from both the incoherent and coherent 
%parts of the spectra, i.e., from all energy scales.

Reducing the doping further, i.e., approaching the $\delta \to 0$ limit,
enhances the asymmetry of the quasiparticle splitted band until
the $\omega > 0$ side is nearly suppressed (spectra D). Significantly,
the Fermi energy gets within the now clean small gap between
the splitted band (due to its weak intensity, the peak in the $\omega >0$ side
can be barely distinguished so is indicated by a double arrow head line 
in the spectra D) \cite{note}.
The system has now become an AF insulator, but qualitatively different
from the large gap Mott state E. Interestingly, the
staggered moment of this very lightly doped state
has almost reached the saturation value, so the small gap value is
not given by $2Um$. Finally, when the doping vanishes,  the strongly
correlated state with low frequency features is no longer sustainable and
the spectra undergoes a rapid redistribution of spectral weight to
acquire the line shape of E (and rigidly shifted by $\mu$ as 
discussed before).

We now study the {\em crossover} from the  Mott insulator to the  strongly correlated
metal induced by  doping in  
 the higher  temperature PM phase.
Previous studies of the  fully frustrated model, which has no AF order, showed that
  at low $T$, the width of the
quasiparticle peak is controlled by the
doping $\delta$ \cite{prb,moeller,jarrell,rmp}.  Since the 
renormalized Fermi energy $\epsilon^*_F(\delta) \sim \delta D$,
it is expected that below the coherence temperature, i.e., 
when $T < \epsilon^*_F(\delta)$ the
system should be a Fermi liquid, while
for  $T > \epsilon^*_F(\delta)$, 
the system should become an incoherent metal, i.e., 
a Mott insulator with 
thermal excitations in the gap.  
%Naively, one  would expect  that
%this scenario  holds for the unfrustrated model in the PM phase at intermediate temperatures.
However, as we shall see,  the crossover
in the intermediate $T$ regime investigated here 
occurs rather differently from this naive expectation.
%In Fig.\ref{fig3}, we present results for the same value $U=3.125$
%and two higher temperatures $T=0.1$ and
%$0.125$.

\begin{figure}
\centerline{\includegraphics[scale=0.5,angle=270]{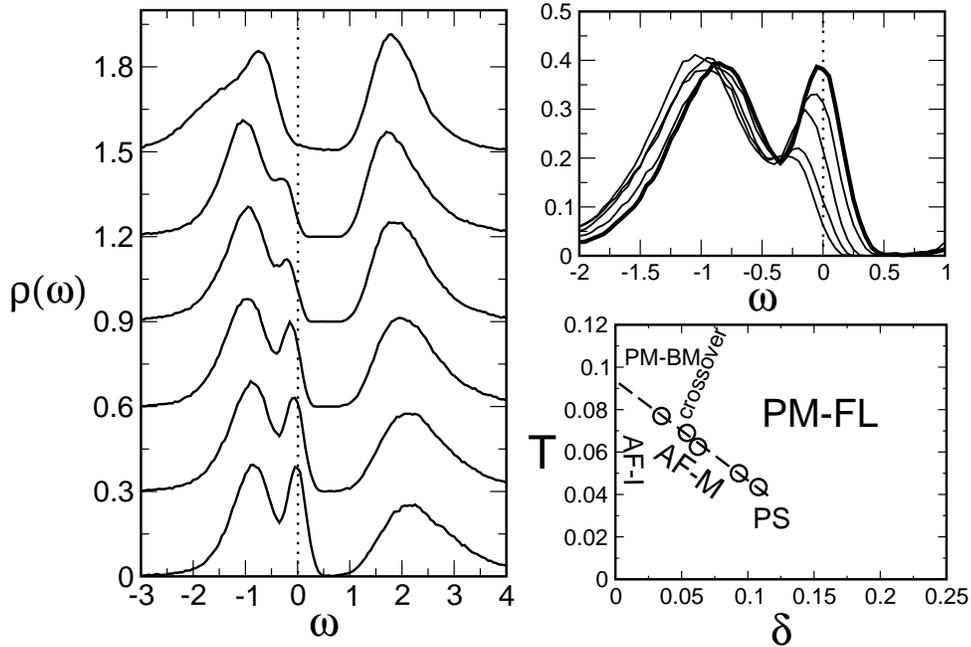}}
%\centerline{\includegraphics[angle=270]{fig2.ps}}
\caption{Left panel: the local DOS $\rho(\omega)$ for $U=3.125$ and
$T= 0.1$. 
The values of the chemical potential are $-\mu$ = 0.4, 0.5, 0.55,
0.65, 0.73 and 0.8, that correspond to $\delta \approx$
0.003, 0.0076, 0.0114, 0.022, 0.038 and 0.055 respectively (top to bottom). 
Top right: the low energy part of the same results 
(-$\mu = 0.4$ is not included and -$\mu = 0.8$ is in thick line).
Bottom right: Schematic phase diagram.
The circles denote our results for $T_N(\delta)$
and are consistent with those of Ref.\onlinecite{freericks}
obtained by computing the susceptibility. 
%FL is for Fermi liquid,
%BM for bad metal, M for metal, I for insulator and PS for phase separation.
}
\label{fig2}
\end{figure}

In the left panel of
Fig.\ref{fig2}  we plot the evolution of the DOS for a series of very low
dopings.
The bottom most curve shows that sufficient doping  results in   a heavy Fermi liquid
metal with a narrow quasiparticle centered at the Fermi energy.
At the other end, the top most curve  shows a virtually insulating
Mott state with negligible DOS at the Fermi energy and a large charge gap ($\sim U$).
The doping of this state is extremely small ($\approx 0.3$\%) and the high temperature
does not allow for a narrow coherent quasiparticle band.
%As expected, the bottom most curve shows 
%the three peak structure of the correlated
%metal. As we  decrease the doping $\delta$,
%the system remains within the paramagnetic phase  due to
%thermal and quantum fluctuations. The most
%salient feature  of the DOS is the loss of spectral 
%weight of the central quasiparticle peak
%which culminates in
%the top most curve ($\delta \approx 0.003$), 
%where the quasiparticle peak vanishes and the
%system becomes a paramagnetic Mott insulator.
The interesting aspect of the crossover between these states
is the anomalous evolution of the quasiparticle peak, 
shown in detail in the top right panel of Fig.\ref{fig2}.
There are two features worth noticing: 
in contrast to the scenario  described earlier 
%for the frustrated model at low $T$,
the quasiparticle peak persists down to very low values
of doping $\delta \sim 0.01$,  in a regime where 
$T >> \delta D$. 
The second unexpected feature is that the position
of quasiparticle peak shifts away from the  Fermi energy. 
The asymmetry that the peak develops around $\omega=0$
can be physically understood as due to the relative suppression
of particle propagation with  respect to hole propagation
in a lightly hole doped Mott insulator (or viceversa).
This anomalous behavior of the quasiparticle peak leads us
to  term  the state in the crossover regime as a ``bad metal'' (BM).
Interestingly, the asymmetry around $\omega=0$ was 
already observed in the low frequency features of the AF state. 
While its origin  in the AF state might have not seemed obvious,
it appears clearer under the new light shed by the
results in the PM state.
It  is perhaps worth mentioning that the anomalous shift
of the quasiparticle peak off the Fermi energy
is reminiscent of the phenomenology of the pseudo-gap state 
observed in photoemission studies of underdoped cuprates \cite{rmp_lee}.
Our results for the evolution of the DOS are summarized in 
the  phase diagram of the Fig.\ref{fig2}.

\begin{figure}
  \centerline{\includegraphics[scale=0.5,angle=270]{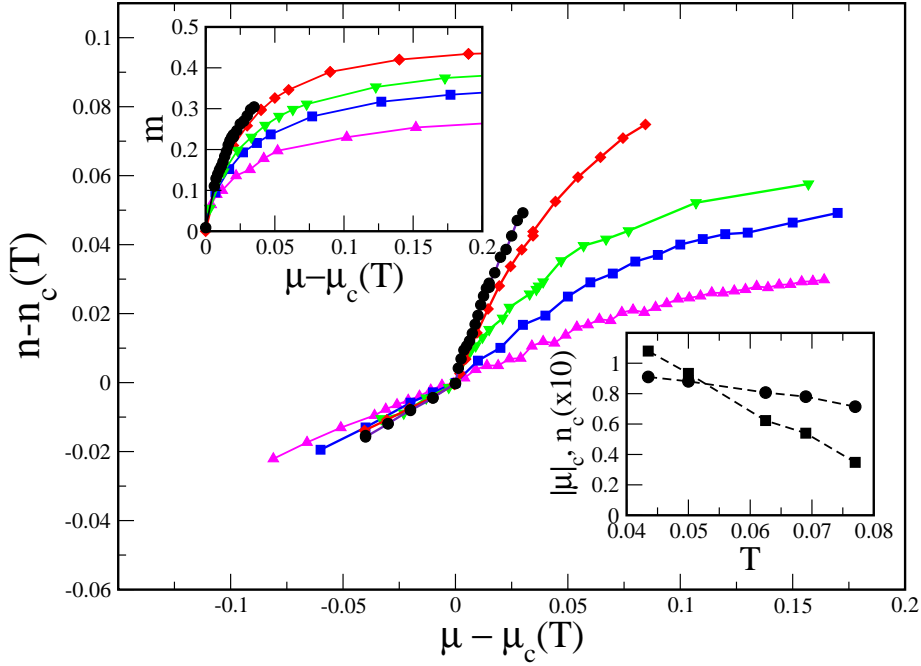}}
\caption{Ocupation $n-n_c(T)$ as a function of the shifted
chemical potential $\mu -\mu_c(T)$ from QMC at $U=3.125$
and different $T$ values.
$T$= 0.043 (black circles), 0.05 (red diamonds),
0.0625 (green down triangles), 0.069 (blue squares) and 0.077 (purple
up triangles). Error bars are about the size of the symbols. Top inset:
$m$ vs. $\mu -\mu_c(T)$. Bottom inset: $|\mu_c|$ (circles)
and $n_c$ (squares) as a function of $T$.
}
\label{fig3}
\end{figure}

Finally, we  consider the critical behavior of the occupation number
$n=1-\delta$ and the staggered magnetization $m$ at the AF/PM boundary.
%Further insight can be obtained from the behavior of the $\delta$
%vs $\mu$ at various temperatures. 
The QMC results are shown in Fig.\ref{fig3}.
The most striking feature is that the slope $ \partial n/ \partial \mu$
which  determines  the  electronic {\em charge} compressibility, shows
a dramatic enhancement on the AF side of the transition line
which corresponds to the AF metal state that we described before.
More significantly, the enhancement rapidly increases as $T$ is lowered.
For instance, reducing $T$ by 50\% yields a 7 fold increase in the 
compressibility. 
Our results are not inconsistent with a divergence at a low finite temperature,
but due to strong numerical fluctuations at the transition boundary
obtaining reliable data at lower $T$ becomes impractical.  Such a divergence
might be a signal of an instability to phase separation.
In any case, this striking behavior is expected to have clear experimental
consequences promoting phase separation (PS), 
anomalies in sound velocity propagation \cite{fournier} and
stuctural or electronic instabilities \cite{hassan} among others.
Remarkably, while the compressibility shows this anomalous behavior, 
the staggered magnetization in contrast displays the expected critical behavior at the
transition. We find $m \propto (\mu-\mu_c)^{1/2}$ for all $T$.

In conclusion, we study the behavior of a doped Mott-Hubbard insulator
within the framework of DMFT. Various interesting regimes were found as the 
model is doped away from both the half-filled antiferromagnetic
and paramagnetic states. In particular, the 
study of the DOS allowed us to identify
a strongly correlated AF metal state and a PM ``bad metal'' one.
The crossover through a ``bad metal'' state was identified.
Along the AF/PM critical line a compressibility anomaly was observed
with a dramatic enhancement at lower temperatures.
The relevance of our results for the interpretation of various
experiments was pointed out. Though a calculation of 
the optical conductivity in these regimes would be clearly interesting, it
is technically challenging and is beyond the scope of the present paper.
Interesting perspectives are opened by this work such as
the investigation of  superconducting  instabilities  and 
 similar features in more realistic multiorbital 
models.

We thank G. Kotliar and M. Gabay for useful comments. 
We acknowledge support from CONICET (PEI6360), ANPCyT PICT03-11609
and UBACyT.

%\begin{figure}
%  \centerline{\includegraphics[scale=0.4,angle=270]{fig4.ps}}
%\caption{DOS for $U=3.125$ and
%$T$=0.1 (top) and 0.125 (bottom). The values of the chemical potential 
%are $-\mu$=0.XXXXXX, and the last one is the curve in thick line.}
%\label{DOSQMCP2}
%\end{figure}

\date{\today}

\begin{thebibliography}{10}
\bibitem{note}
This result was confirmed by an exact enumeration calculation
of the Green's function followed by a Pade analytic continuation.
\end{thebibliography}
\end{document}